# Hard disk drive as a magnetomechanical logic device

Vladimir L. Safonov, *Senior Member, IEEE*

*Abstract*—We consider the conditions how two binary numbers can be superimposed on the same track with the use of different recording magnetic fields. As a result the average magnetization of longitudinal medium along the track can have three states: *-M*, *0* and *+M*. Possibility to perform logic operations with these states is considered. We demonstrate OR, AND, XOR and NOT operations and discuss a modification of a recording device.

## I. INTRODUCTION

HARD DISK drive is usually used as just a storage device in personal computers. Here we consider a possibility to perform logic operations on a hard disk drive. This can be done on a magnetic recording medium in which there is a possibility to superimpose binary information. Such media consist of magnetic particles with i) different anisotropy fields and/or with ii) different anisotropy axes orientations. The first is, for example, an oriented (e.g., perpendicular) medium with a distribution of anisotropy fields. The second is typical for a longitudinal medium where the anisotropy axes of particles are distributed randomly.

The elementary logic operation consists of three steps: 1) Applying a strong magnetic field, one writes the first binary number *A* as the transitions between different average magnetization states (*−M* and *+M*, all previous information is erased). 2) Using smaller magnetic field that reverses just a part of magnetic particles, we write the second binary number *B*. Doing so we obtain transitions between 0 and *M*, *-M* and 0, and *−M* and *M*. 3) The reading sensor detects the peaks of magnetization variation between the uniformly magnetized regions on the recording magnetic surface and gives result of the logic operation OR, AND, XOR or NOT. Using the results of XOR and AND operations, we can construct an adder.

## II. SUPERIMPOSITION OF THE BINARY INFORMATION

Let us demonstrate how the binary information can be superimposed on the recording track. We shall consider a longitudinal medium containing identical particles with the uni-axial anisotropy field $H_K$ and randomly distributed anisotropy axes in the disk plane. If the external magnetic field of the writing head $H_1$ is greater than $H_K$, all particles obtain magnetization components oriented along this field (see, Fig.1).

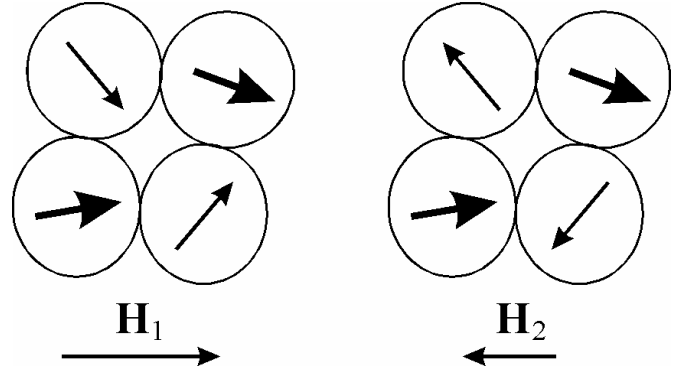

Fig.1. Example of superimposing. The bold and thin arrows denote the magnetizations of type 1 and type 2 particles, respectively.

The absolute average magnetization of the medium is *M*. Changing the direction of field $H_1$, we erase all the previous binary information and obtain a written pattern with *–M* and *+M* states and transitions between these states (see, Fig. 2a).

Then to the same recording track we apply the magnetic field $H_2$, which is less than $H_K$. Nothing is changed if this field is oriented in the same direction as averaged magnetization. But if the field $H_2$ has the opposite direction (see, Fig.1), the magnetic particles can be separated in two types. Type 1 particles do not change their states. The anisotropy axes of these particles are deviated from the track direction by the angle $\phi$: $|\phi| \leq |\phi_0|$. And type 2 particles with $|\phi_0| \leq |\phi| \leq \pi/2$ reverse to the opposite direction. Here the angle $\phi_0$ is defined by the equation: $H_2 = H_K \cos \phi_0$. The resulting magnetization is a sum of magnetizations of "type 1" and "type 2" particles. It is equal to zero if $H_2 = H_K \sqrt{3}/2 \simeq 0.87 H_K$. The binary information is written on the type 2 particles in this case (see, Fig. 2b).

So, applying fields $H_1$ and $H_2$ independently in the same and opposite directions, we superimpose the binary information for two binary numbers *A* and *B*. As the result of two subsequent field applications, we have the average magnetization of the recording medium (see, Fig. 2c).

## III. LOGIC OPERATIONS

In order to develop logic operations, we need to represent the binary numbers in a doubled form changing 0 by 00 and 1 by 11. For example, the binary number 11010 should be represented in the following form (11)(11)(00)(11)(00). Here the parentheses are shown for simplicity of representation. Technically this representation can be made with the doubling of the clock frequency.

In Fig. 2a, b the binary information is represented in the forms of "pulses" (11) or (**I**) of magnetization $M$ and $M/2$, respectively, relative to the $-M$ and $-M/2$ states. The height of pulses could be $M$ or $2M$. The binary numbers are: $A$=(11)(00)(11)(00) or $A$=**I0I0** (**I** denotes 11 and **0** denotes 00), $B$=(11)(00)(00)(11) or $B$=**I00I**.

### A. OR and AND logic operations

Figure 2 illustrates the OR operation. For a simple peak detector that does not distinguish the height of transitions ($M$ and $2M$), from the result of superimposing in Fig. 2c we obtain the following sequence of binary information: (11)(00)(11)(11), or $C$ = **I0II** = (**I0I0**)OR(**I00I**) = ($A$)OR($B$).

For a peak detector (or an extra peak detector) that detects the only largest ($2M$) transitions between $-M$ and $M$, from the superimposing we can have AND logical operation: $D$ = **I000** = (**I0I0**)AND(**I00I**) = ($A$)AND($B$).

### B. XOR operation

We can introduce a concept of a "negative" (**-I**) pulse. In this case the pulse is considered down relative to $M$ or $M/2$ states. In Fig. 3b we see an example how the binary number B=**I00I** is shown in a negative form $-B$=**-I00-I**. In other words, $-B$ is a reversed magnetization profile of $B$ relative to horizontal axis $M$=$0$. It is easy to check that ($B$)OR($-B$)=**0000**.

Figure 3 illustrates the XOR operation. Superimposing the first number $A$ = **I0I0** and the second number $B$ (in the form of $-B$ = **-I00-I**), we obtain the resulting magnetization profile $C$ = **00I-I**. So far as the sign of the pulse in $C$ does not matter for a peak detector, $C$ = **00II** = (**I0I0**)XOR(**I00I**) = ($A$)XOR($B$).

Actually, XOR logic operation can also be obtained from Fig. 2c if one assumes that a peak detector can separate signals with $2M$ and $M$ variations. As it was mentioned above, for $2M$ variation (left peak in Fig. 2c) we have AND operation (all other signals are zeros). For $M$ variation (two smaller peaks in the right side of Fig. 2c) we have XOR operation (with the assumption that for $2M$ variation signal is zero for this specific peak detector).

### C. NOT operation

Consider the following operation: (**0**)OR(**-I**) = **-I**. We can neglect the sign (a peak detector detects just the magnetization variation) and write (**0**)OR(**-I**) =**I**. In addition, as it was pointed out above, we have (**I**)OR(**-I**)=0. Thus, OR operation with **-I** inverts the initial binary number, and NOT operation is just OR operation for any binary number and corresponding sequence of **-I**. For example, NOT(**I0I0**) = (**I0I0**)OR(-**I-I-I-I**) = **0I0I**, where the sign (-**I**) in the result is omitted.

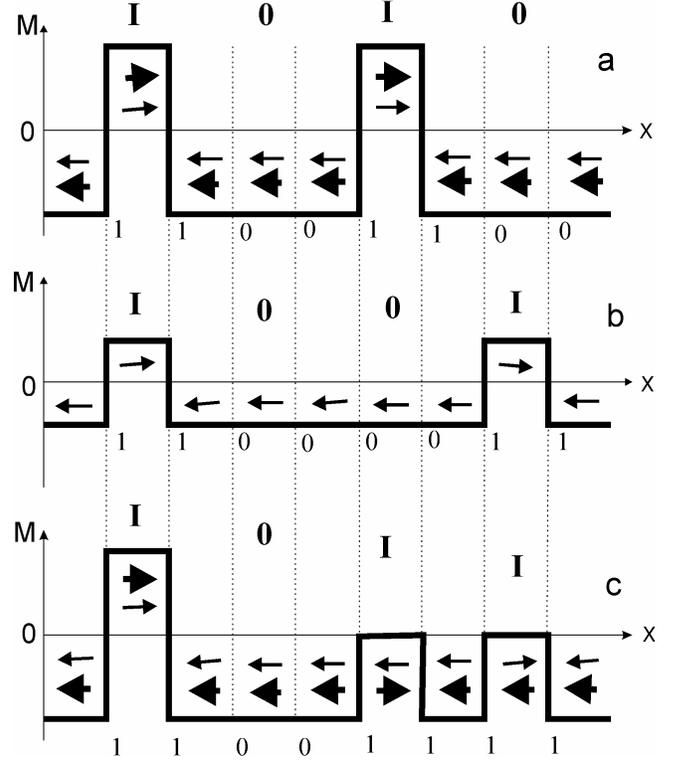

Fig.2. Magnetization profile (M) along the track (x) written with a) $H_1$ and b) $H_2$. c) The result of superimposition. The vertical dotted lines separate different bits. The ones below each profile denote transitions and zeros indicate the absence of transitions. The bold and thin arrows denote the magnetizations corresponding to type 1 and type 2 particles, respectively.

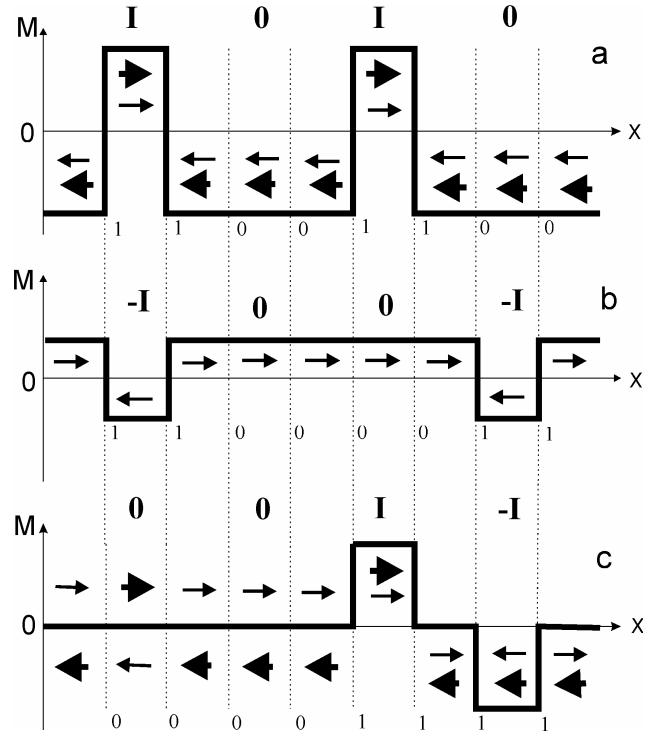

Fig.3. Magnetization profile (M) along the track (x) written with a) $H_1$ and b) $H_2$. c) The result of superimposition.

All other logic operations, such as NAND, NOR and XNOR can be constructed from the OR, XOR, AND and NOT operations: $(A)$NAND$(B)$ = NOT$[(A)$AND$(B)]$, $(A)$NOR$(B)$ = NOT$[(A)$OR$(B)]$, $(A)$XNOR$(B)$ = NOT$[(A)$XOR$(B)]$.

### D. Adder

The results of $A1 = (A)$XOR$(B)$ and $C1 = (A)$AND$(B)$ logic operations can be used to construct a binary adder. We represent $B1$ = LeftShift$(C1)$, where LeftShift() denotes the left shifted $C1$ by one register. Then one finds $A2 = (A1)$XOR$(B1)$ and $C2 = (A1)$AND$(B1)$ and again $B2$ = LeftShift$(C2)$. All procedure repeated $n$ times until $Cn$ is equal to zero in all registers. The final result is $An$. The other arithmetic operations can be developed in a similar way. Thus we can organize a complete logic and numeric processor.

### E. Block diagram

A block diagram of the logic device is similar to hard disk drive conventional block diagram (see, e.g., [1]) with some modifications.

The input sequentially gives binary numbers and the instruction to perform logic operations OR, XOR, AND or NOT. For example, A=1010 and B=1001.

Then the logic gates and decoder transform the binary numbers into doubled form (0 to 00, 1 to 11), for example, $A$=(11)(00)(11)(00)=**I0I0** and $B$=(11)(00)(00)(11)=**I00I**. The logic gates can control also the peak detector operations.

Modulation creates the voltage signal corresponding to A (and then B) for the recording head.

Recording system consists of recording head, recording medium and reading sensor. Changing the current of the recording head, we apply the magnetic fields $\pm H_1$ and $\pm H_2$ to the recording medium. The reading sensor detects the magnetization transitions.

The peak detector reads the signal of the reading sensor and detects the peaks of magnetization variation. A simple peak detector does not distinguish the small variations (the transitions between $-M/2$ and $M/2$, $-M$ and 0, or 0 and $M$) and large variations (the transitions between $-M$ and $M$). An advanced peak detector can detect the only large transitions (between $-M$ and $M$) and makes it possible to perform AND operation (see, Fig. 2c).

Finally, a decoder transforms **I**=11 to 1 and **0**=00 to 0 and sends it to output.

## IV. DISCUSSION

All the above description regarding the superimposed binary information and logic operations is applicable also for an oriented (perpendicular) recording media. For example, one can use the perpendicular medium with two types of particles with the uniaxial anisotropy fields $H_{K1} > H_{K2}$, $H_1 \geq H_{K1}$ and $H_{K2} \leq H_2 < H_{K1}$. The absolute values of average magnetizations of both types of particles should be equal to each other ($M/2$ and $M/2$, respectively). The arrows in Figs. 2, 3 must be vertical in this case.

So far as the recording time necessary for logic operation is very small, the recording medium can consist of particles with small $H_K$. This will reduce the heating of the medium. The particles can also have smaller size, which is important to reduce the noise. To solve the problem of synchronization of bits at superimposition we can increase the linear size of a pulse (e.g., introducing extra zeros for coding, 11 becomes 101, 1001, and so on).

Logic operation is performed during 3 full rotations of a hard disk drive with a standard head. However, it is possible to develop a "tandem head" consisting of 1) first writing head that creates $\pm H_1$, 2) second writing head that creates $\pm H_2$, and 3) reading sensor. Such a tandem head will write, superimpose and read the binary information along the track continuously.

Logic operations on a hard disk drive can be used both for independent data processing and as a supplemental tool for a conventional semiconductor processor. The main advantage is that the binary numbers can be very long ones (correspond to a full turn of a hard disk drive). Similar logic operations can be also developed for a magnetic tape and any other recording media with superimposing.

Performing various logic and arithmetic operations on a hard disk, we can develop a magnetomechanical digital processor. The speed of this processor is defined by the number of revolutions per second ($\sim$100 sec$^{-1}$) multiplied by the bit length of the track ($\sim$10$^6$ bit).

### ACKNOWLEDGMENT

The Author thanks A. K. Khitrin for valuable discussions.